\documentstyle[12pt]{article}
\begin{document}
\newcommand{\nd}[1]{/\hspace{-0.5em} #1}
\begin{titlepage}
\begin{flushright}
{\bf January 2000} \\ 
SWAT-249  \\
UW/PT 00-02 \\ 
hep-th/0001103 \\
\end{flushright}
\begin{centering}
\vspace{.2in}
{\large {\bf Softly-Broken ${\cal N}=4$ 
Supersymmetry in the Large-$N$ Limit }}\\
\vspace{.4in}
 Nick Dorey \\
\vspace{.4in}
Department of Physics, University of Wales Swansea \\
Singleton Park, Swansea, SA2 8PP, UK\\
\vspace{.2in}
and \\ 
\vspace{.2in}
 S. Prem Kumar \\
\vspace{.4in}
Department of Physics, University of Washington, Box 351560   \\
Seattle, Washington 98195-1560, USA\\
\vspace{.4in}
{\bf Abstract} \\
\end{centering}
We calculate the exact values of the holomorphic observables of 
${\cal N}=4$ supersymmetric $SU(N)$
Yang-Mills theory deformed by mass terms which preserve ${\cal N}=1$
SUSY. These include the chiral condensates in each massive vacuum 
of the theory as well as the central charge which determines the
tension of BPS saturated domain walls interpolating between these vacua. 
Several unexpected features emerge in the large-$N$ limit, including 
anomalous modular properties under an $SL(2,Z)$ duality group which acts
on a complexification of the 't Hooft coupling $\lambda=g^{2}N/4\pi$. 
We discuss our results in the context of the
AdS/CFT correspondence.

\end{titlepage}
\section{Introduction}
\paragraph{} 
The AdS/CFT correspondence \cite{mal} makes non-trivial
predictions for the large-$N$ behaviour of four-dimensional gauge
theories. Although the correspondence was initially applied to 
the ${\cal N}=4$ supersymmetric Yang-Mills (SYM) theory at the
conformal point in its moduli space, subsequent
developments have led to IIB supergravity duals for gauge 
theories with less supersymmetry obtained by perturbing the 
${\cal N}=4$ theory with relevant operators. 
These perturbations can lead to new conformal field
theories in the IR \cite{conformal}, as well as massive theories 
in confining \cite{confine,gir}, Coulomb \cite{coulomb} or 
Higgs phases. A general feature of each of these cases is that the 
supergravity approximation is valid only for large
values of the 't Hooft coupling, $\lambda=g^{2}N/4\pi\gg 1$. 
To go beyond this regime, one would have to evaluate stringy 
corrections to the supergravity limit exactly, which is not feasible 
at present. On the other hand, conventional gauge theory calculations 
are usually only possible for $\lambda \ll 1$, or not at all in the 
case of the confining phase. One way of making progress around this impasse is 
to identify special 
quantities on either side of the correspondence whose dependence on 
$\lambda$ is highly constrained by 
supersymmetry and can be determined exactly. 
The main examples of this type in the ${\cal
N}=4$ theory are the two- and three-point correlation functions of chiral
primary operators \cite{three} which are known to be completely independent of
$\lambda$. Instanton contributions to four-point functions (and
higher) \cite{dhkmv} seem to be similarly protected although the origin of
$\lambda$-independence in this case is less well understood 
(see however \cite{green}) . 
\paragraph{}
In this letter we will consider the massive deformations of ${\cal N}=4$
SYM which preserve ${\cal N}=1$ supersymmetry. The resulting theory 
exhibits a variety of supersymmetric ground-states in 
confining, Coulomb and Higgs phases. As is usual for theories with 
${\cal N}=1$ supersymmetry, there is a special class of quantities 
which depend holomorphically on the parameters \cite{seib}. 
In particular, these
include the vacuum expectation values of the lowest components of 
chiral superfields or `chiral condensates' for short. In
four-dimensions, the ${\cal N}=1$ supersymmetry
algebra admits a central charge which is also a holomorphic function
of the parameters \cite{shifman}. The resulting Bogomol'nyi bound leads to an 
exact formula for the tension of BPS-saturated domain walls
\footnote{We will not, however, address the much more complicated
problem of checking that these BPS domain walls are actually present
in the theory.}. In recent work \cite{me}, one of the authors derived the exact
superpotential of the mass-deformed ${\cal N}=4$ theory. In the following we
will use this and related results to determine the chiral condensates in each
massive supersymmetric vacuum and the tension of BPS-saturated domain walls
interpolating between each pair of vacua 
(at least when $N$ is a prime number). The resulting formulae
transform covariantly under the S-duality group of the ${\cal N}=4$
theory. In particular, the central charge transforms with modular
weight two up to permutations of the vacua. 
\paragraph{}
The main application of the field theory results described above
is to investigate the large-$N$ limit of the theory in its confining
phase. For
$\lambda\ll 1$, the scalar fields of the ${\cal N}=4$ theory (and their ${\cal
N}=1$ superpartners) decouple and the theory reduces to pure ${\cal N}=1$
supersymmetric Yang-Mills (SYM) theory. 
This theory is expected to be qualitatively
similar to the large-$N$ limit of QCD, exhibiting asymptotic freedom in the UV
and confinement in the IR. 
On the other hand, for $\lambda \gg 1$, the theory
potentially has a description in terms of IIB supergravity on a background
which is asymptotically $AdS_{5}\times S^{5}$. Indeed a supergravity dual 
has recently been proposed by Girardello et
al \cite{gir}. These authors show that the dual description reproduces
several qualitative properties of the confining phase such as gluino
condensation, a mass gap and confinement itself. However, as
the dual geometry is singular, it is not clear whether the supergravity
approximation should be quantitatively reliable, even when the 't
Hooft coupling is large. 
\paragraph{}
Our exact formulae interpolate between these two regimes
and we discuss their implications for the dual string theory. 
Several surprising features emerge in the large-$N$ limit. 
The chiral condensates have quasi-modular properties under an $SL(2,Z)$ 
duality group which acts on a complexification of the 't Hooft
coupling. Taken at face value, this corresponds to an anomalous 
T-duality of the dual string theory which inverts the radius of the
asymptotic geometry. We use this unexpected property to
determine the leading behaviour of the holomorphic quantities in the 
$\lambda\rightarrow \infty$ limit explicitly. 
We find that the vacuum
structure of the large-$\lambda$ theory in its confining phase is 
qualitatively different from 
that of ${\cal N}=1$ SYM. In the large-$N$ limit, 
the latter theory has a continuous degeneracy of vacua lying on a 
circle centered at the origin in the complex ${\cal S}$-plane (here
${\cal S}$ denotes the gluino condensate). For $\lambda\gg 1$, 
we find instead an infinite but discrete set of points 
in the ${\cal S}$-plane, all lying on a finite segment of a 
single line through the origin. This behaviour reflects an 
interesting non-analyticity of the exact formulae near the point 
$\lambda=\infty$, $N=\infty$ in parameter space. 
\paragraph{}
If the theory does have a supergravity dual for large $\lambda$, 
then the resulting series expansion around $\lambda=\infty$ 
represents a series of 
stringy corrections to the supergravity limit. 
However, we find that the series contains 
exponentially supressed terms which have no obvious semiclassical 
interpretation as worldsheet instantons \cite{green}. By analogy 
with similar phenomena in quantum field theory, we take this as a 
hint that the corresponding world-sheet description may be 
intrinsically strongly coupled. 
We also discuss the large-$N$ 
behaviour of the theory in its unique Higgs phase vacuum. 
The large-$N$ scaling of the holomorphic quantities in this 
vacuum raises a puzzle. In particular, we find that the tension of a
BPS-saturated domain wall interpolating between vacua in the 
Higgs and confinement 
phases scales like $N^{4}$. This is hard to interpret in the context of a
dual string theory with string coupling $g_{st}\sim 1/N$. We briefly 
discuss a possible resolution of this puzzle.    
\paragraph{}
We begin by considering ${\cal N}=4$ supersymmetric Yang-Mills (SYM) 
theory with gauge group $SU(N)$ in four dimensions. 
In terms of ${\cal N}=1$ 
superfields this theory contains a gauge multiplet $V$ as 
well as three chiral multiplets $\Phi_{i}$, $i=1,2,3$ in the adjoint 
representation of $SU(N)$. Soft breaking to ${\cal N}=1$ is accomplished
by introducing masses $m_{i}$ for each chiral superfield $\Phi_{i}$. 
Including these terms the classical superpotential is\footnote{The
normalization of the superpotential differs from \cite{me} in two
respects. Firstly, as is standard when discussing the large-$N$ limit,
we have rescaled the fields so that each term in the classical
Lagrangian scales like $N$. Secondly, the factor of $2\sqrt{2}$
ensures that our conventions agree with those of Seiberg and Witten
\cite{SW2} for gauge group $SU(2)$.}, 
\begin{equation}
W=N{\rm Tr}_{N}\left(2\sqrt{2}\Phi_{1}[\Phi_{2},\Phi_{3}]+ 
m_{1}\Phi_{1}^{2}+ m_{2}\Phi_{2}^{2}+m_{3}\Phi_{3}^{2}\right).
\label{spmass}
\end{equation} 
Following \cite{VW}, we will now analyse vacuum
structure of the theory. 
It is convenient to rescale the chiral superfields 
as $\Phi_{1}=i\sqrt{m_{2}m_{3}} X/\sqrt{2}$,  
$\Phi_{2}=i\sqrt{m_{3}m_{1}} Y/\sqrt{2}$, 
$\Phi_{3}=-i\sqrt{m_{1}m_{2}} Z/\sqrt{2}$. Up to an overall
normalization the superpotential then becomes, 
\begin{equation}
W=N{\rm Tr}_{N}\left(\frac{i}{2}(X^{2}+Y^{2}+Z^{2})-X[Y,Z] \right).
\label{spxyz}
\end{equation} 
To find supersymmetric ground states, we must impose the F-term equations 
which come from varying $W$ with respect to $X$, $Y$ and $Z$. 
The first equation is $iX=[Y,Z]$ and the other two are generated by 
cyclic permutation of the three superfields. A supersymmetric vacuum 
is therefore described by three $N\times N$ matrices which obey the 
commutation relations of an $SU(2)$ Lie algebra. However, we still 
have to impose the corresponding D-term equations and mod out by $SU(N)$ gauge 
transformations. As usual these two steps can be combined by dividing out 
the action of the complexified gauge group $SL(N,C)$ on $X$, $Y$ and $Z$. 
In fact, up to an $SL(N,C)$ gauge transformation, there is exactly one 
solution of the $SU(2)$ commutation relations in terms of $N\times N$ matrices 
for each $N$-dimensional representation of $SU(2)$ \cite{VW}. 
\paragraph{}
We will first consider the unique irreducible representation of dimension $N$. 
Fixing the $SL(N,C)$ gauge symmetry, we may set 
$X$, $Y$ and $Z$ equal to the standard generators $J_{X}$,
$J_{Y}$ and $J_{Z}$ of this representation. Thus we have 
$Z=J_{Z}={\rm diag}(m,m-1,\ldots,1-m,-m)$ with $m=(N-1)/2$.  
This choice of VEVs for the scalar fields completely breaks 
the $SU(N)$ gauge symmetry and thus we have a single supersymmetric 
vacuum state in the Higgs phase. One may wonder whether the 
classical analysis leading to this conclusion is 
reliable. This depends on whether we can choose the parameters so that 
the theory in this vacuum is weakly
coupled. At energies much larger than $|m_{i}|$,
the mass terms are irrelevant and we have ${\cal N}=4$ supersymmetric
Yang-Mills theory. In this theory the $\beta$-function vanishes exactly and
the complexified gauge coupling $\tau=4\pi i/g^{2}+\theta/2\pi$ does
not run. On the other hand, as the gauge symmetry is completely
broken, the classical vacuum has a mass gap and thus the gauge 
coupling does not run for mass scales much less than $|m_{i}|$ either.
Thus, as long as the 't Hooft coupling $\lambda=g^{2}N/4\pi$ is much
less than one the theory is weakly coupled at all energy scales as required. 
\paragraph{}
In this vacuum we may easily evaluate the classical values 
of the chiral condensates, $u_{l}=\langle{\rm Tr}_{N}
\Phi_{3}^{l}\rangle$. We find that  
\begin{eqnarray}
u_{2k} &= &-\frac{m^{k}_{1}m^{k}_{2}}{2^{k-1}} \sum_{l=1}^{r} l^{2k}
\;\;\;\mbox{for}\;\; N=2r+1; \nonumber \\
& = & -\frac{m_{1}^{k}m_{2}^{k}}{2^{3k-1}} \sum_{l=1}^{r} (2l-1)^{2k}
\;\;\;\mbox{for}\;\; N=2r. 
\label{evenu2a}
\end{eqnarray}
Evaluating these sums explicitly for $k=1$ we find, 
$u_{2}=m_{1}m_{2}N(1-N^{2})/24$ for all $N$. 
Note also that the $u_{2k+1}$ vanish for each
value of $N$. Expanding (\ref{evenu2a}) in the large-$N$ limit we 
obtain, 
\begin{equation}
u_{2k}=-\frac{1}{2^{3k}(2k+1)}m^{k}_{1}m^{k}_{2}
N^{2k+1}\left(1+
O\left(\frac{1}{N^{2}}\right)\right),
\label{u2ka}
\end{equation}
For later convenience we 
introduce the normalized condensates $\tilde{u}_{2k}=u_{2k}/N^{2k+1}$,
which approach constant values in the large-$N$ limit. 
\paragraph{}
We may also anticipate the form of quantum corrections to these results. As
discussed above, the theory is weakly-coupled near $\tau=i\infty$ and
should therefore have a semiclassical expansion around this point. 
Perturbative corrections correspond to powers of $1/{\rm
Im}(\tau)$ while instanton and anti-instanton contributions are weighted by 
factors of $q=\exp(2\pi i \tau)$ and $\bar{q}$ respectively. 
However, $u_{2k}$ is the expectation value of the lowest component of
a chiral superfield and must therefore be holomorphic 
in $\tau$. Hence we expect $u_{2k}$ to have an instanton expansion
in integer powers of $q$ with no perturbative corrections. Contributions from 
anti-instantons or $I\bar{I}$ pairs are also ruled out by holomorphy. 
As, $q \sim \exp(-2\pi N/\lambda)$, each of these instanton contributions is
exponentially supressed in the large-$N$ limit. Thus we find that 
the classical results (\ref{evenu2a}) for the condensates $u_{2k}$ in
the Higgs phase vacuum, are exact at each finite order in the $1/N$
expansion. Thus (\ref{evenu2a}) yields definite predictions for the 
condensates in the limit $\lambda\rightarrow \infty$ 
which we discuss in the context of the AdS/CFT correspondence below. 
\paragraph{}
The trivial representation of $SU(2)$ yields another classical
supersymmetric vacuum. In this case we have $\Phi_{i}=0$ for $i=1,2,3$
and the gauge group is completely unbroken. At energy scales much less
than $|m_{i}|$, the three hypermultiplets decouple and the low-energy theory
reduces to ${\cal N}=1$ SUSY Yang-Mills with gauge group $SU(N)$. 
This theory has a negative $\beta$-function and the gauge
coupling becomes strong at the dynamical scale 
$\Lambda=(m_{1}m_{2}m_{3})^{\frac{1}{3}}q^{\frac{1}{3N}}$ 
leading to confinement and the generation of a mass gap. Further,  
${\cal N}=1$ SYM has a non-anomalous $Z_{2N}$ global symmetry 
which is spontaneously broken to $Z_{2}$ by gluino condensation 
giving $N$ supersymmetric vacua \cite{AK}. The theory is believed
to have BPS-saturated domain walls which interpolate between each pair
of vacua \cite{shifman}. The chiral condensates
will depend holomorphically on the complexified gauge coupling $\tau$
and, as above, this rules out any perturbative contributions to these
quantities. However, an important difference between the Higgs and confinement
phases is that the latter is strongly-coupled in the IR irrespective 
of how small the gauge coupling of the UV theory is. Hence, the form 
of non-perturbative contributions is not constrained by semiclassical 
considerations. In particular, fractional powers of $q$, which
cannot be attributed to any finite action classical field
configuration, may occur. 
\paragraph{}
In summary, we have found a total of $N+1$ 
massive supersymmetric vacua. The theory is in the Higgs phase in 
one of these and the confining phase in the remainder. Of course, we 
have only considered vacua corresponding to the irreducible and the 
trivial representations of $SU(2)$ and there are many intermediate 
possibilities. It turns out that as long as $N$ is prime, all the
remaining vacua correspond to a Coulomb phase without a mass gap
\cite{VW}. We
will not consider these massless vacuum states here. If $N$ has
non-trivial divisors, then there are also additional confining vacua 
where a non-abelian proper subgroup of $SU(N)$ remains unbroken. In fact the
total number of massive vacua is equal to the 
sum of the divisors of $N$. To avoid complication we will 
restrict our attention to those confining vacua where the full $SU(N)$ gauge
group remains unbroken (corresponding to the trivial represenation of $SU(2)$).
These are precisely the $N$ confining vacua that appear when $N$ is prime.
Many of the results that we present below can however, be generalised easily to
include the remaining vacua as well.
\paragraph{}
In \cite{me}, one of the authors derived the exact superpotential for
the theory considered above via compactification on $R^{3}\times
S^1$. The results presented in \cite{me} are independent of the radius of
compactification and apply equally to the theory on $R^{4}$.  
For the present purposes, we will only require the value of the 
superpotential in each vacuum as a function of the masses $m_{i}$ and the
complexified gauge coupling $\tau$. 
Labelling the $N$ confining vacua with an integer 
$m=0,1,\ldots, N-1$, the superpotential in each of these vacua is given as,    
\begin{equation}
W=\frac{N^{3}}{24} m_{1}m_{2}m_{3} \left[g_{m}(\tau,N) +
A(\tau,N)\right]\label{exact1}
\end{equation}
while in the Higgs vacuum we have, 
\begin{equation}
W=\frac{N^{3}}{24} m_{1}m_{2}m_{3} \left[h(\tau,N) +
A(\tau,N)\right]\label{exact2}
\end{equation}  
with, 
\begin{eqnarray}
g_{m}(\tau,N) & = & E_{2}(\tau)-
\frac{1}{N}E_{2}\left(\frac{\tau+m}{N}\right)\\\nonumber\\
h(\tau,N) & = & E_{2}(\tau)-NE_{2}(N\tau).
\label{gh}
\end{eqnarray}
Here $A(\tau,N)$ is an undetermined holomorphic
function of $\tau$ with a weak coupling expansion of the form, 
\begin{equation}
A(\tau,N)= \alpha_{0}+\sum_{k=1}^{\infty} \alpha_{k}q^{k} 
\end{equation}
which will be discussed further below. 
\paragraph{}
In the above, 
$E_{2}(\tau)$ is the regulated second Eisenstein series \cite{KOB}, 
\begin{equation}
E_{2}(\tau)=\frac{3}{\pi^{2}}
\sum_{n=-\infty}^{+\infty} {\sum_{m=-\infty}^{+\infty}}'
 \frac{1}{(m+n\tau)^{2}}.
\label{eis}
\end{equation}
The prime on the $m$ summation means `omit the $m=0$ term when
$n=0$'. In the weak coupling limit\footnote{In the following, we will 
use the term `weak-coupling' to describe the expansion of our exact
results for large ${\rm Im}(\tau)$ even though, as explained above,
the theory in the confining phase is not weakly coupled in this regime.}, 
${\rm Im}(\tau)\rightarrow \infty$, $E_{2}(\tau)$ can be expanded as, 
\begin{equation}
E_{2}(\tau)= 1-24\sum_{k=1}^{\infty} c_{k}q^{k} 
\label{eisexp}
\end{equation}
with $c_{k}=\sum_{d|k}d$. Hence we see that the weak coupling
expansion of the superpotential in the Higgs vacuum contains only 
integral powers of $q$ as anticipated above. In contrast, the
expansion of the superpotential in each of the confining vacua 
is a power series in $q^{\frac{1}{N}}$. 
These contributions formally
correspond to fractional instantons or `merons'. Note that 
we have $q^{\frac{1}{N}}\sim \exp(-2\pi/\lambda)$ in the 't Hooft
limit. This means that unlike the
effects of conventional instantons which are exponentially supressed
at large-$N$, fractional instanton contributions are
independent of $N$ in this limit.  
Of course, there are
no corresponding solutions of the classical field equations 
with fractional toplogical charge 
for a gauge theory on $R^{4}$. The presence of 
fractional powers of $q$ simply reflects the fact that the theory 
in the confining phase is intrinsically strongly coupled and thus 
semiclassical reasoning is invalid. Interestingly however, for the
theory compactified on $R^3 \times S^{1}$, a semiclassical
interpretation of merons does emerge (at least when the radius of $S^1$ is
small): they are precisely magnetic monopoles which contribute as
three-dimensional instantons after compactification \cite{LY,HKMS} 
(see also \cite{VB} and references therein). 
\paragraph{}
The superpotential determines the value of the condensates 
$u_{2}=\langle {\rm Tr}_{N}\Phi_{3}^{2} \rangle$ and ${\cal S}=\langle 
{\rm Tr}_{N} W_{\alpha}W^{\alpha}\rangle$ via the standard relations, 
\begin{eqnarray}
u_{2}=\frac{1}{N} \frac{\partial W}{\partial m_{3}}\;;& \qquad{}
\qquad{} & 
{\cal S}=-8\pi i\frac{\partial W}{\partial \tau}.
\label{rel} 
\end{eqnarray}
Using the methods of \cite{me}, one may also calculate the 
exact values of the chiral condensates $u_{2k}$ with $k>1$ in each
vacuum of the theory. Some explicit results along these lines are
given in the Appendix. 
\paragraph{}
From the first relation in (\ref{rel}), we see that 
the unknown function $A(\tau,N)$ results in a nonperturbative ambiguity  
in the condensate $u_{2}$.  As we discuss below, 
this coincides with the ambiguity \cite{DKM4} in the definition of 
the corresponding quantity in ${\cal N}=2$ SYM with one adjoint
hypermultiplet. The first coefficient, 
$\alpha_{0}$, may be fixed by comparison with the 
classical analysis given above. In particular, we require that $u_{2}$ 
vanishes in the weak-coupling limit ${\rm Im}(\tau)\rightarrow \infty$ 
in each confining vacuum. This requires $\alpha_{0}=(1-N)/N$. If we now 
substitute this value into the corresponding formula for the Higgs
branch vacuum we find $u_{2}\rightarrow m_{1}m_{2}N(1-N^{2})/24$ as 
${\rm Im}(\tau)\rightarrow \infty$ in precise agreement with 
Eq (\ref{evenu2a}). 
The rest of the coefficients $\alpha_{k}$, $k>0$, remain unknown
except in the case of gauge group $SU(2)$ where $\alpha_{1}$ was
determined in \cite{DKM4} by an explicit instanton calculation.     
\paragraph{}
The above formulae have interesting properties under the S-duality
group of ${\cal N}=4$ SUSY Yang-Mills. Specifically for instance, when $N$ is
prime, under the 
electric-magnetic duality transformation $S: \tau\rightarrow -1/\tau$ we have, 
\begin{eqnarray}
g_{0}\left(-\frac{1}{\tau},N \right) = \tau^{2}h(\tau,N)\;;& \qquad \qquad
& g_{m}\left(-\frac{1}{\tau}, N\right) = \tau^{2}g_{p}(\tau,N)\;;
\nonumber \\ 
h\left(-\frac{1}{\tau}, N\right) = \tau^{2}g_{0}(\tau,N)\;;& & 
\label{stransfmn}
\end{eqnarray}
where the second equality applies for $m=1,\ldots,N-1$, with $p(m,N)$ 
denoting the least positive integer satisfying $\;pm=-1\;{\rm
mod}\;N$. Under the other generator of the modular group, 
$T: \tau\rightarrow \tau + 1$ we have, 
\begin{eqnarray}
g_{m}(\tau+1,N) = g_{m+1}(\tau,N)\;;& \qquad \qquad
& g_{N-1}(\tau+1,N) = g_{0}(\tau,N)\;;\nonumber \\
h(\tau+1,N) = h(\tau,N).
\label{ttransfmn}
\end{eqnarray}
Hence the set of functions $\{g_{0},g_{1},\ldots,g_{N-1},h\}$
transform as modular forms of weight two modulo permutations. In the
$SU(2)$ case, the three functions in question  are just the 
well-known quasi-modular forms, $e_{1}(\tau)$, $e_{2}(\tau)$ and
$e_{3}(\tau)$. The presence of the unknown function
$A(\tau,N)$ implies that these modular properties are not 
inherited by the superpotential itself. In order to recover modular 
covariant results we have to work with the shifted superpotential 
$\tilde{W}=W-N^{3}m_{1}m_{2}m_{3}A(\tau,N)/24$. Fortunately, as we
will see below, this shift does not affect the domain wall tensions of
the theory as these only depend on the differences of the
value of the superpotential in different vacua. 
\paragraph{}
The fact that $SL(2,Z)$ transformations permute the different vacua has
a nice explanation in terms of the corresponding theory with ${\cal
N}=2$ supersymmetry which is obtained by setting $m_{1}=m_{2}=M$ and $m_{3}=0$
\cite{SW2,DW}.
In this case the theory has a Coulomb branch parameterized by the
moduli $u_{l}$ introduced above, which is determined by a
hyperelliptic curve. On special singular submanifolds of the Coulomb branch 
the curve degenerates producing massless BPS states. If we now
reintroduce a non-zero value for $m_{3}$, thereby breaking ${\cal N}=2$
supersymmetry down to an ${\cal N}=1$ subalgebra, the massless BPS
states at each singular point condense giving a massive supersymmetric
vacuum. The values of the condensates $u_{l}$ in each vacuum of the 
${\cal N}=1$ theory are determined by the location of the corresponding
singular point on the Coulomb branch of the ${\cal N}=2$ theory.   
The natural action of the S-duality group on the 
BPS spectrum is then inherited by the singular points on the Coulomb
branch and thus by the corresponding ${\cal N}=1$ vacua. 
In the case of gauge group $SU(2)$, the 
ambiguity arising from the unknown function $A(\tau,N)$, 
corresponds to the mismatch between the parameter $\tilde{u}$
appearing in the Seiberg-Witten curve \cite{SW2}, which
transforms with modular weight two, and the 
physical quantity $u=\langle {\rm Tr}_{2} \Phi^2 \rangle$ which is not
modular covariant \cite{DKM4}.     
\paragraph{}
In general the theory may contain BPS-saturated domain walls which
interpolate between each pair of supersymmetric vacua. To determine  
whether BPS-saturated domain walls exist, we would need more
information about the theory beyond 
the holomorphic quantities calculated in this paper. 
However, if these states are present, the exact
formula for their tension is $T=|\Delta W|$ where 
the central charge $\Delta W$ is the difference between the value 
of the superpotential in each of the two vacua \cite{shifman}. 
Thus we may find a BPS saturated domain wall interpolating
between the Higgs vacuum and each confining vacuum with, 
\begin{equation}
\Delta W_{m}=\frac{N^{4}}{24} m_{1}m_{2}m_{3} \left[E_{2}(N\tau)
-\frac{1}{N^{2}}E_{2}\left(\frac{\tau+m}{N}\right) \right] 
\label{higgsm}
\end{equation}
and between each pair of confining vacua with, 
\begin{equation}
\Delta W_{m,n}=\frac{N^{2}}{24} m_{1}m_{2}m_{3} 
\left[E_{2}\left(\frac{\tau+m}{N}\right)-E_{2}
\left(\frac{\tau+n}{N}\right)\right].
\label{dwmn}
\end{equation}
The fact that these expressions scale differently in the large-$N$
limit will be significant in what follows. 
As $SL(2,Z)$ transformations permute the supersymmetric vacua, they
also permute the domain walls which interpolate between each pair of
vacua. In particular, the domain wall tensions $T=|\Delta W|$ 
transform with (non-holomorphic) modular 
weight $(1,1)$ up to permutations. Given the existence of one type 
of BPS-domain wall in the theory, this modular behaviour suggests 
that all the others should be present. 
\paragraph{}
The theory in each of the confining vacua flows to pure ${\cal N}=1$
supersymmetric Yang-Mills theory in the IR. More precisely, to
decouple the extra degrees of freedom of the ${\cal N}=4$ theory we 
must must ensure that the dynamical scale $\Lambda=
(m_{1}m_{2}m_{3})^{\frac{1}{3}}q^{\frac{1}{3N}}$ is much less than each
of the chiral multiplet masses. This will be the case if the 't Hooft coupling
$\lambda=g^{2}N/4\pi$ is much less than one. In this regime it is easy
to check that our formulae reproduce the standard results for ${\cal
N}=1$ SUSY Yang-Mills. 
In particular, the gluino condensate in each of
the confining vacua is \cite{AK}, 
\begin{equation}
{\cal S}(m)=16\pi^2Nm_{1}m_{2}m_{3}
\exp\left(-\frac{8\pi^{2}}{g^{2}N}\right)
\exp\left(i\frac{2\pi m}{N}+i\frac{\theta}{N}\right)=
16\pi^2N\Lambda^3\exp\left(\frac{2\pi i m}{N}\right)
\label{n=1s}
\end{equation}
for $m=0,1,\ldots,N-1$. The leading correction to the above formula 
is of order $\exp(-16\pi^{2}/g^{2}N)$ which is small provided 
$\lambda\ll 1$. Note that 
the $\theta$ parameter has been absorbed in the definition of the
dynamical scale $\Lambda$ reflecting the presence of an anomalous
$U(1)_{R}$ symmetry which appears in the decoupling limit. 
The confining vacua lie at the vertices
of a regular $N$-gon in the complex 
${\cal S}$-plane. This reflects the spontanteous breaking of a
non-anomalous $Z_{2N}$ subgroup of $U(1)_{R}$ down to $Z_{2}$. 
Importantly, neither $U(1)_{R}$ or its anomaly-free subgroup are 
exact symmetries of the mass-deformed ${\cal N}=4$ theory. 
We also find the standard formula for the tension of BPS saturated 
domain walls interpolating between pairs of confining vacua 
\cite{shifman}, 
\begin{equation}
T_{m,n}=|\Delta W_{m,n}|=
16\pi^2N^{2}\Lambda^{3}|1-e^{\frac{2\pi i}{N}(m-n)}|
\label{n=1tension}
\end{equation}
Note that the Higgs
phase vacuum of the mass-deformed ${\cal N}=4$ theory effectively
decouples in this limit. In particular the ratio of the 
tension of a domain wall 
interpolating between the Higgs and confining phases to that of one
interpolating between two confining vacua diverges as $\lambda
\rightarrow 0$.
\paragraph{}
In \cite{witqcd}, Witten made the interesting proposal that the domain
wall of ${\cal N}=1$ supersymmetric Yang-Mills theory can act as
Dirichlet brane for the confining string. 
The modular properties of the supersymmetric vacua and the domain
walls which may interpolate between them described
above suggests a generalization of this phenomenon which may occur in
the present model. As explained above, each supersymmetric vacuum 
is associated with the condensation of a set of BPS states of the 
${\cal N}=2$ theory of \cite{DW}. This in turn leads 
to the confinement of other BPS states of the ${\cal N}=2$ theory. 
The simplest example is that condensation of electrically charged 
elementary fields in the Higgs phase vacuum leads to the confinement 
of magnetic monopoles and vice versa for the confining vacuum with 
$m=0$. The confinement of each kind of charge requires a 
string which carries the corresponding flux. 
This suggests that we should find an $SL(2,Z)$ multiplet of confining 
strings each of which can end on one kind of domain wall. 
The situation is similar to the
properties of strings and fivebranes in Type IIB string theory. 
There one finds an $SL(2,Z)$ multiplet of strings and of the
fivebranes on which they may terminate. Each fivebrane 
acts as a Dirichlet brane for one type of string. It is tempting to
believe that this similarity with the IIB theory is more than a
coincidence and can be explained in terms of the AdS/CFT
correspondence. 
\paragraph{}
In the rest of the paper we will discuss the physics of the 't Hooft limit 
$N\rightarrow\infty, g^2\rightarrow 0$ with $\lambda=g^{2}N/4\pi$ held fixed. 
On general
grounds \cite{TH},  
we expect the large-$N$ gauge theory to have a dual description 
as a string theory with string coupling $g_{st}$ proportional to
$1/N$. As the theory in the confining phase reduces to 
${\cal N}=1$ supersymmetric Yang-Mills for $\lambda \ll 1$, we will 
begin by reviewing the large-$N$ limit of this theory in as 
discussed in \cite{witqcd}. 
First note that the  $\Lambda$-parameter defined above 
depends only on the 't
Hooft coupling and not on $N$ itself. Thus the dynamical mass scale of
the theory is held fixed as $N\rightarrow\infty$. The theory 
is believed to contain an infinite tower colour 
singlet excitations or `glueballs' with masses of order $N^{0}$ in
units of $\Lambda$. One may also argue that each term in the effective
action for these degrees of freedom scales like $N^{2}$, reflecting 
the fact that the 
glueballs are stable and weakly interacting in the large-$N$ limit. 
This is consistent with the idea that the large-$N$ gauge theory 
is equivalent to a weakly coupled string theory. The leading order
terms in the effective action are then identified as a genus zero 
contribution of the string. As above, the confining vacua 
lie at the vertices of a regular $N$-gon in the complex ${\cal S}$-plane. As $N\rightarrow \infty$, the $N$-gon approaches a circle and 
we have a continuous degeneracy of vacua. Note, however, that 
this does {\em not} imply that a massless
goldstone mode appears in the large-$N$ limit\footnote{The 
explanation of this point is related to the
fact that the $U(1)$ problem of the $\eta$ mass in QCD can be
resolved within the context of the $1/N$ expansion \cite{sigma}}.  
One may also try to
identify the string theory states which correspond to the BPS 
saturated domain walls of the theory. Naively the tension $T_{m,n}$ 
of a domain wall given in (\ref{n=1tension}) above scales like
$1/g_{st}^{2}$ which is typical of a finite energy classical field
configuration such as a soliton. However, this is only true if 
$m-n\sim N$. For domain walls
interpolating between adjacent vacua (or any pair of vacua with
$m-n\sim N^{0}$) we find instead $T_{m,m+1}\sim 1/g_{st}$. 
This scaling is typical for a Dirichlet brane 
rather than a soliton. This observation motivates Witten's proposal 
that the domain walls are D-branes for the confining string. 
The above features of ${\cal N}=1$ SYM can also be understood by 
realizing this theory on the worldvolume of an M-theory fivebrane 
\cite{witqcd}.  
\paragraph{}
As the ${\cal N}=1$ theory discussed in this paper can be realized as
a relevant deformation of ${\cal N}=4$ SUSY Yang-Mills, the dual
string theory should 
be determined by the AdS/CFT correspondence. 
In particular, the large-$N$ gauge theory should have a dual
description in terms of IIB string theory, with string coupling 
$g_{st}\sim 1/N$, on a spacetime manifold 
which is asymptotically $AdS_{5}\times S^{5}$. 
The 't Hooft coupling of the gauge theory 
corresponds to the ratio $L^{4}/4\pi \alpha'^{2}$ on the IIB side, 
where $L$ is the common radius of the $AdS_{5}$ and $S^{5}$ 
factors of the asymptotic geometry and $\sqrt{\alpha'}$ is the 
string length-scale. In the case of the undeformed
${\cal N}=4$ theory, the IIB background is exactly 
$AdS_{5}\times S^{5}$ 
and IIB string theory can be approximated by 
IIB supergravity as long as $\lambda\gg 1$. However, 
for the mass-deformed 
theory, the relevant IIB geometry may be singular, in which case 
the supergravity approximation is not valid even for large 
$\lambda$ \cite{gir}. In principle, our field theory 
results yield predictions for the dual string theory. 
However, for a meaningful comparison we must identify 
the observables on the IIB side which correspond 
to the holomorphic gauge theory quantities 
calculated in this paper. As discussed further below, 
chiral condensates correspond to certain moduli 
of the IIB geometry \cite{krauss, klebanov}. In addition, 
domain walls interpolating between vacua should correspond 
to IIB brane configurations which preserve two supercharges although 
this has not been investigated previously. 
The double expansion 
of the holomorphic quantities 
in inverse powers\footnote{The expansions also
contain  non-perturbative terms in each of these parameters.} of $N$ 
and $\lambda$ then provides information about 
string-loop and worldsheet effects respectively. 
Obviously, most 
of these predictions would be very hard to test given our 
current ignorance about string theory in Ramond-Ramond (RR) 
backgrounds. However, the main point of the 
analysis is to obtain explicit results for the holomorphic 
observables in the limit $\lambda\rightarrow\infty$ and 
investigate the implications for a dual description in terms of 
IIB supergravity like that proposed in \cite{gir}. 
\paragraph{}
We will now consider the large-$N$ behaviour of the mass-deformed 
${\cal N}=4$ theory in the $N$ confining vacua labelled by the integer
$m=0,1,\ldots,N-1$. Note that $m$ always appears in
the formulae via the combination $\kappa=(m+\theta/2\pi)/N$. As we
vary $m$ and $\theta$, $\kappa$ can take any value in the interval 
$[0,1]$. Initially we will think of $\kappa$ as a continuous 
variable of order $N^{0}$. 
In fact $\kappa$ itself only enters in a complex combination
with the 't Hooft coupling, 
\begin{equation}
\rho=\frac{\tau+m}{N}=\frac{i}{\lambda}+\kappa
\label{rho}
\end{equation}
We begin by implementing the 't Hooft limit in the standard way, 
taking the $N\rightarrow \infty$ limit with $\lambda$ 
(and thus $\rho$) held fixed. 
The gluino condensate in each confining 
vacuum is given by, 
\begin{equation}
{\cal S}(m)=\frac{N\pi i}{3} m_{1}m_{2}m_{3}\left[ E'_{2}(\rho)+ 
O\left(\exp\left(\frac{-2\pi
N}{\lambda}\right)\right)\right]
\label{gcexact}
\end{equation} 
Note that the 
ambiguity in the definition of the condensate corresponding to 
the unknown function $A(\tau,N)$ is exponentially supressed 
in the large-$N$ limit. 
\paragraph{}
Interestingly, the leading order term 
has quasi-modular properties in the complexfied 't Hooft coupling
$\rho=i/\lambda+\kappa$. In particular, the second Eisenstein series 
$E_{2}(\rho)$ is invariant under $\rho\rightarrow
\rho+1$ but has the following anomalous transformation law under the 
other generator of $SL(2,Z)$,  
\begin{equation}
E_{2}(\rho)=\frac{1}{\rho^{2}}E_{2}\left(-\frac{1}{\rho}\right)-
\frac{6}{\pi i\rho}
\label{anomaly}
\end{equation}
The violation of modular invariance is mild in the sense that it can
be recovered by modifying $E_{2}(\rho)$ by a non-holomorphic 
correction \cite{vafa}. 
The behaviour of the gluino condensate under a general modular 
transformation $\rho\rightarrow \tilde{\rho}=(a\rho+b)/(c\rho+d)$, 
 with integers $a$,$b$,$c$,$d$ satisfying $ad-bc=1$,  
is governed by the formula,  
\begin{equation}
E'_{2}(\rho)=\frac{1}{(c\rho+d)^{4}}E'_{2}\left(\tilde{\rho}\right)-
\frac{2c}{(c\rho+d)^{3}}E_{2}\left(\tilde{\rho}\right)+
\frac{6}{\pi i}\frac{c^{2}}{(c\rho+d)^{2}}
\label{anomaly2}
\end{equation}
This quasi-modular behaviour in $\rho$ originates in the modular
properties of the function $g_{m}(\tau,N)$ appearing
in the exact superpotential (\ref{exact1}) under S-duality.  
In fact the anomalous term reflects the fact that 
S-duality transformations do not commute with the 't Hooft limit. 
However, although it has its origin in S-duality 
on the gauge theory side, the modular group acting on $\rho$ 
does not correspond to the usual S-duality group of 
IIB string theory. 
If we set $\kappa=0$ 
(ie set $\theta=0$ and focus on the confining
vacuum with $m=0$), then the transformation 
$\rho\rightarrow -1/\rho$ simply inverts the 't Hooft
coupling $\lambda \rightarrow 1/\lambda$. As above 
the 't Hooft coupling $\lambda$ is identified with the
ratio $L^{4}/4\pi \alpha'^{2}$ on the 
IIB side. Therefore, in terms of string theory parameters, 
the corresponding transformation is an 
anomalous T-duality which inverts the radius of the 
asymptotic geometry! 
\paragraph{}
The anomalous transformation law (\ref{anomaly2}) 
allows us to obtain the expansion of the gluino condensate in the 
supergravity limit\footnote{For convenience we will refer to the
limit $\lambda\rightarrow \infty$ as the supergravity limit even
though a supergravity description may not be valid there.}    
$\lambda\rightarrow\infty$.  
Explicitly (with $\kappa=0$) we have, 
\begin{equation}
E'_{2}(\rho)=E'_{2}\left(\frac{i}{\lambda}\right)=
 -2i\lambda^{3} +\frac{6i}{\pi }\lambda^{2} -48 i\lambda^{3}
\sum_{k=1}^{\infty}
(\pi k\lambda -1)c_{k}\exp(-2\pi k \lambda) 
\label{eisexp3}
\end{equation}
with $c_{k}=\sum_{d|k}d$ as above. 
With the standard identification of parameters given above,  
(\ref{eisexp3}) represents a series of stringy  
corrections to the supergravity limit, 
with expansion parameter $\alpha'^{2}/L^{4}=1/\lambda$. 
In a more 
conventional string theory compactification without background 
RR fields, one would hope to identify exponentially 
suppressed terms like 
those appearing in (\ref{eisexp3}) as the saddle-point contribution 
of an instanton obtained by wrapping an appropriate BPS state
around a non-contractible cycle in spacetime. In fact, as the string 
coupling does not enter explicitly, the most natural possibility 
would be worldsheet instantons coming from wrapping of the 
IIB string itself \cite{green}. 
This is not promising in the present context for a number of reasons. 
A worldsheet instanton would typically have an action proportional to 
$\sqrt {\lambda} \sim L^{2}/\alpha'$ rather than $\lambda$ itself. 
In fact, on general grounds, one would also expect the subleading 
terms in (\ref{eisexp3}) to be down by powers of $1/\sqrt{\lambda}$
rather than $1/\lambda$. 
In any case there are no topologically non-trivial cycles of the right
dimension in $AdS_{5}\times S^{5}$, nor are any apparant in the
deformation of this geometry considered in \cite{gir}. 
Thus there is no obvious semiclassical interpretation for 
the exponentially suppressed terms appearing in (\ref{eisexp3}).
Because very little is known about worldsheet effects for string
theory in RR backgrounds, it is hard to draw any definite conclusion
from this. However it is perhaps worth making a comparison with 
a similar phenomena we met in a different context above: the occurence
of fractional instanton contributions in the confining vacua of the
four dimensional gauge theory. By analogy with that case, 
our result may simply suggest that the relevant world-sheet
description is intrinsically strongly coupled.
\paragraph{}
In our discussion of ${\cal N}=1$ SUSY Yang-Mills theory at large-$N$
we found a continuous degeneracy of confining 
vacua lying on a circle centred at the origin of the 
complex ${\cal S}$-plane. This analysis applies to the 
confining vacua of the mass-deformed ${\cal N}=4$ theory for $\lambda\ll 1$.     
From the point of view of the AdS/CFT correspondence, an obvious
question is what happens to this picture for $\lambda \gg 1$. 
The large-$\lambda$ expansion of the gluino condensate given in
Eq. (\ref{eisexp3}) is specific to  
the confining vacuum with $m=0$ at $\theta=0$. 
The generalization of this formula 
to the remaining vacua is surprisingly subtle. 
\footnote{For any finite $\lambda$, the 
vacua of the large-$N$ theory will fill a closed curve in 
the ${\cal S}$-plane, parameterized by $\kappa\in[0,1]$, 
which generalizes the circle of the 
${\cal N}=1$ SYM case. However as $\lambda$ increases the curve 
becomes increasingly irregular and it does not approach a 
well-defined limit as $\lambda\rightarrow\infty$.}. 
One sensible way to study the large-$N$ limit in the confining vacua with
$m\neq 0$ is to consider various
sub-sequences $\{m(N)\}$ such that $\lim_{N\rightarrow\infty} {m(N)/N}$ converges to
some rational number $p/q$ \footnote{Other ways of taking the large-$N$,
large-${\lambda}$ limit -- e.g. taking $\lambda$ to infinity keeping $N$ fixed
also leaves us with the rational points only with the gluino condensate
having the same leading behaviour as obtained below.}. Here $p$
and $q$ are
mutually prime.
It is then convenient to perform a modular transformation of the form, 
\begin{equation}
\rho\rightarrow \tilde{\rho}=\frac{a\rho+b}{q\rho-p}
\label{rhotilde}
\end{equation}
using (\ref{anomaly2}), 
where the modular condition is satisfied by choosing $a$ and $b$ 
such that $ap+bq=-1$. In the large-$\lambda$ limit, the leading behaviour of
the gluino condensate in the vacuum labelled by a rational number $p/q$ is,

\begin{equation}
{\cal S}=\frac{2\pi}{3} Nm_{1}m_{2}m_{3}\frac{\lambda^{3}}{q^{2}} 
\label{sm}
\end{equation}
Note that for given $q$, the gluino condensate has the same leading behaviour
for every $p$ less than $q$ such that $p$ and $q$ are coprime.
Thus, in the large-$N$ limit, we find an infinite tower of vacua located at the points, 
\begin{equation}
{\cal S}(j)=\frac{2\pi}{3} N m_{1}m_{2}m_{3}\frac{\lambda^{3}}{j^{2}} 
\label{sm1}
\end{equation} 
in the complex ${\cal S}$-plane where $j$ runs over the positive 
integers. 
Further, the above formula is only valid for $\theta=0$. 
Changing $\theta$ continuously from $0$ to $2\pi$ induces a complicated 
rearrangement of the vacua. 
Although the vacua have a point of accumulation at
${\cal S}=0$ there is no trace of the continuous vacuum degeneracy which
appears in the large-$N$ limit at small $\lambda$. Note also that 
the discrete vacuum degeneracy for large $\lambda$ is 
associated with the magnitude of the
condensate only: the phase of ${\cal S}$ is the same in each vacuum. 
In contrast, for small $\lambda$, the magnitude of the 
condensate is the same in each vacuum because of the spontaneously 
broken $Z_{2N}$ symmetry which appears in this limit. 
\paragraph{}
A concrete application of the field theory results described
above is to test the dual description in terms of IIB supergravity 
proposed in \cite{gir}. 
In particular, the chiral condensates in a given vacuum correspond 
to the amplitudes of certain normalizable zero modes in the expansion 
of the asymptotic supergravity fields around the $AdS_{5}\times 
S^{5}$ background \cite{krauss,klebanov}. 
Indeed, the authors of \cite{gir} used this correspondence 
to demonstrate the existence of a non-zero gluino condensate starting
from their supergravity solution. For the configuration
considered in \cite{gir}, the magnitude of the gluino condensate 
is essentially a free parameter which 
arises as a constant of integration when solving the 
five-dimensional field equations. This corresponds to 
a continuous degeneracy of supersymmetric vacua on the field theory 
side of the correspondence which disagrees with the results 
presented above. Presumably this means that stringy effects 
remain important even at large 't Hooft coupling and must have the
effect of lifting the continuous vacuum degeneracy.    
However, the supergravity description does reproduce one 
qualitative feature of the field theory results\footnote{The authors 
would like to thank Alberto Zaffaroni for this
observation.}. Specifically, the vacuum degeneracy implied by 
the supergravity solution of \cite{gir} is associated with the 
magnitude of the condensate rather than its phase 
in agreement with the large-$\lambda$ formula (\ref{sm1}). 
\paragraph{} 
So far we have only considered the large-$N$ limit of the theory 
in its confining phase. However, in the spirit of the 
AdS/CFT correspondence, we might expect that each vacuum state of the
theory has a dual description in terms of weakly-coupled 
IIB string theory on a manifold which is asymptotically 
$AdS_{5}\times S^{5}$. Similarly, one expects that 
the BPS-saturated domain
wall interpolating between each pair of vacua should have a
counterpart in the IIB Hilbert space. 
However, if we try to extend the above 
discussion to the unique vacuum of the theory in the 
Higgs phase, we immediately confront a problem.     
According to equation (\ref{higgsm}), the exact tension of the 
domain wall which interpolates between the Higgs and confinement 
phases grows like $N^{4}\sim 1/g^{4}_{st}$ which is 
unacceptable in a weakly-coupled string theory\footnote{Of course one
way to avoid this conclusion is simply to suppose that the offending 
BPS state does not exist.}. 
At this point we may simply decide that our assumptions 
were too strong and that the theory in the 
Higgs phase vacuum cannot be described in terms 
of a IIB background which is asymptotically
$AdS_{5}\times S^{5}$. However we 
will also explore the consequences of another more speculative 
resolution. Firstly, note that the appropriate normalization 
of the domain wall tension for a comparison with the 
tension of a BPS state on the IIB side of the AdS/CFT 
correspondence is not obvious. In particular it is possible 
that extra factors of $g_{st}\sim 1/N$ could appear. 
It is therefore possible that the $N^{4}$ 
scaling of the exact superpotential in the 
Higgs phase vacuum actually corresponds to the leading order 
in string perturbation theory on the IIB side. 
Of course this 
means that we should interpret the superpotential in the 
confining phase, which scales like $N^{2}$, 
as a genus one effect. With this interpretation, there is a 
continuous degeneracy of confining vacua at string tree-level 
which is lifted by a one-loop effect in string perturbation theory. 
Clearly this clashes with the 
standard discussion of the large-$N$ limit of ${\cal N}=1$ SYM 
as reviewed above. In particular the domain walls interpolating
between confining vacua with $m-n\sim N$, would now correspond to
states with tension of order $g_{st}^{0}$ on the IIB side which 
could not be interpreted as classical solitons. The interpretation 
of domain walls interpolating between adjacent vacua as D-branes 
would also be spoilt. 
\paragraph{}
Finally we will consider the large-$\lambda$ limit of
the chiral condensates in the Higgs phase vacuum. 
The classical formulae (\ref{evenu2a}) for the 
$u_{2k}$ are exact to all finite orders in the $1/N$ expansion 
(ie the only corrections are non-perturbative in $1/N$).  
In particular, $u_{2k}$ scales like $N^{2k+1}$ in the large-$N$ 
limit with a coefficient which does not depend on $\lambda$. 
As for the gluino condensate, the condensates $u_{2k}$ 
are associated with the amplitudes of certain 
normalizable modes in the expansion of the IIB background around 
$AdS_{5}\times S^{5}$ \cite{krauss,klebanov}. 
The fact that the condensates grow rapidly with $N$ suggests 
that an expansion around $AdS_{5}\times S^{5}$ may not make sense. 
This seems to favour the more conservative resolution of the puzzle
discussed above, namely that the theory in the Higgs vacuum does not
have a sensible large-$N$ limit. However, once again, this conclusion 
depends on our assumptions about the correct normalization of 
the quantities on either side of the correspondence. In particular,  
in the above, we defined the normalized condensates 
$\tilde{u}_{2k}$ which tend to  
the constant values $-m_{1}^{k}m_{2}^{k}/2^{3k}(2k+1)$ in the
large-$N$ limit. An alternative interpretation is that the 
$\tilde{u}_{2k}$ rather than the $u_{2k}$ should be identified 
with the corresponding modes of the supergravity fields in an
expansion around $AdS_{5}\times S^{5}$. With this new 
interpretation, our results suggest that this ground state 
should have a dual description in terms of a classical
solution of the supergravity equations. Unlike the proposed
supergravity dual for the confining vacuum, this solution should be
an isolated one (in otherwords there should be no vacuum degeneracy at
leading order in $1/N$). The corresponding configuration should 
involve non-zero background values for the supergravity fields which
couple to the chiral operators $\tilde{u}_{2k}$ of the gauge theory. 
According to the ideas of \cite{coulomb,krauss,klebanov}, 
the relevant configuration should correspond 
to a non-trivial distribution of D3 branes in
ten dimensions. However there will also be background values for the 
NS-NS antisymmetric tensor field which couple to the bare masses 
of the fermions in each chiral multiplet. The presence 
of these background fields breaks ${\cal N}=4$ supersymmetry 
down to an ${\cal N}=1$ subalgebra (in four-dimensional conventions) 
just as the chiral multiplet 
masses do on the Yang-Mills side of the correspondence. 
Our result for the condensates $\tilde{u}_{2k}$, should then 
translate into a quantitative prediction for the holomorphic moments 
of the D3-brane distribution. 
\paragraph{}
The authors would like to thank Massimo Porrati and 
Alberto Zaffaroni for commenting on a first draft of this paper.  
ND also would like to thank Michael Green and Misha Shifman 
for useful discussions. ND acknowledges the support of TMR 
network grant FMRX-CT96-0012 and of a PPARC advanced research 
fellowship. SPK would like to acknowledge the support 
of DOE grant DE-FG03-96ER40956. 

\section*{Appendix}
In this appendix, we give some exact results for the condensates
$u_{2k}$ for $k>1$, a more detailed discussion of these results will
appear elsewhere \cite{WIP}. In general these quantities will  
suffer from ambiguities like the one discussed in the text for
$u_{2}$. However, we will suppress this fact in the following. 
\paragraph{}
In \cite{me}, one of the authors analysed the theory considered here
via compactification on $R^{3}\times S^{1}$. The main result was an
exact expression for the superpotential on the Coulomb branch of the
three-dimensional effective theory, which coincides with the complexified
potential of the elliptic Calogero-Moser system. This result, 
which can be deduced from holomorphy combined with
various semiclassical considerations, can also be derived from the 
general connection \cite{R, MW, DW} between integrable systems and 
supersymmetric gauge theories. A novel feature discussed in 
\cite{me}, is that each equilibrium configuration of the integrable 
system corresponds to a supersymmetric vacuum of the ${\cal N}=1$ 
theory. The condensates $u_l=\langle {\rm Tr} \Phi_3^l\rangle$ 
in each vacuum are then identified with the constants of motion of the
Calogero-Moser integrable system in a given equilibrium configuration.
The exact values for the chiral condensates in the massive vacua are
implicitly given by, 
\begin{eqnarray}
&&C_{2k}(u)=\frac{(-1)^{2k} m_{1}^{k}m_{2}^{k}}{(2k)!!} 
\sum_{a_1\neq a_2\neq a_3...\neq a_{2k}}{\cal P}(X_{a_1}-X_{a_2})
{\cal P}(X_{a_3}-X_{a_4})... {\cal
P}(X_{a_{2k-1}}-X_{a_{2k}}), \nonumber\\
&&C_{2k+1}(u)=0,
\label{ck}
\end{eqnarray}
where ${\cal P}$ is the Weierstrass function and the $C_k$ are 
Schur polynomials in $u_k$ which are the coefficients of
$t^{-k}$ in $\sum_{k=0} C_k(u) t^{-k} = \exp[-\sum_{s=0}^\infty u_s
/t^{-s}]$. In the Higgs vacuum we have $X^a=2\pi ia/N$, 
($a=1,\ldots,N-1$) 
while in the confining vacuum with $m=0$ discussed in the text 
we have $X^a=2\pi i\tau a/N$, ($a=1,\ldots,N-1$).
\paragraph{}
It is then easy to see that in the 't Hooft large-$N$ limit 
$u_{2k}\sim
m_{2}^{k}m_{1}^{k}N^{2k+1}$ in the Higgs vacuum 
which coincides with 
our earlier weak-coupling analysis (\ref{evenu2a}). 
On the other hand in the confining vacuum we find  
that $u_{2k}\sim m_{1}^{k}m_{2}^{k}N^{k+1}$. 
An explicit example is provided by $u_2$ which by Eq. (\ref{ck}) 
is equal to 
$\frac{m_{1}m_{2}}{2}\sum_{a_1\neq a_2}{\cal P}(X_{a_1}-X_{a_2})$. 
This sum can be evaluated explicitly to obtain the exact results 
for $u_{2}$ given in the text. 
If we consider the appropriately normalised condensates ${\tilde
u}_{2k}=u_{2k}/N^{2k+1}$ , then in the Higgs vacuum they approach their
semiclassical values, while in the confining vacua they vanish in the
large-$N$ limit.

\end{document}